# Theory of Dynamic Stripe Induced Superconductivity

Annette Bussmann-Holder, K. Alex Müller[1], Roman Micnas[2], Helmut Büttner[3], Arndt Simon, Alan R. Bishop[4] and Takeshi Egami[5]

Max-Planck-Institut für Festkörperforschung, Heisenbergstr.1, D-70569 Stuttgart, Germany
[1]Physik Institut Universität Zürich, Winterthurerstr.190, CH-8057 Zürich, Switzerland
[2]Institute of Physics, A. Mickiewicz University, 85 Umultowska St., 61-614 Poznan, Poland
[3]Lehrstuhl für Theoretische Physik I, Universität Bayreuth, D-95440 Bayreuth, Germany
[4]Los Alamos National Laboratory, Theoretical Division, Los Alamos, NM 87545, USA
[5]Department of Materials Science, University of Pennsylvania, LRSM-3231 Walnut Street Philadelphia, PA 19104, USA

Since the recently reported giant isotope effect on T* [1] could be consistently explained within an anharmonic spin-charge-phonon interaction model, we consider here the role played by stripe formation on the superconducting properties within the same model. This is a two-component scenario and we recast its basic elements into a BCS effective Hamiltonian. We find that the stripe formation is vital to high-$T_c$ superconductivity since it provides the glue between the two components to enhance $T_c$ to the unexpectedly large values observed experimentally.

The existence of nanoscale spin – charge phase separation (including „stripes") in cuprates was demonstrated by the observation of the striped phase [2, 3]. While the stripe phase was confirmed directly only in the non – superconducting phase, similar nanoscale phase separation is strongly supported even in the superconducting phase of the cuprates by inelastic neutron scattering and EXAFS measurements [4, 5, 6]. The role of this phase separation for superconductivity is open, and various approaches exist which either question any importance of it to superconductivity [7] or consider it as supporting pair formation [8]. However, up to now there is no detailed microscopic theory which emphazises its role for the unusual high transition temperatures observed in copper oxide superconductors. On the other hand there is growing experimental and theoretical support [9] that all these systems can be understood in terms of a two-component scenario [10]. Yet it remains unclear what these two components are and what is the glue which combines them. We will show in the following that the two components relate to spin and charge type excitations and that their interactions stems from phonons. Thus, HTSC is indeed a remarkable collusion of spin, charge and lattice.

The antiferromagnetic parent compounds of HTSC have provoked models that relate their superconducting properties to antiferromagnetic fluctuations since the observed transition temperatures exceed conventional BCS [11] or Eliashberg [12] theoretical predictions. These approaches [13] neglect effects arising from the lattice since a large Hubbard U is attributed to the copper ion sites which is thought to justify the use of a single band Hubbard model or t-J model or phenomenological antiferromagnetic spin fluctuations [14]. In such approaches the effects of phonons on the superconducting state are usually discarded even though there is broad experimental support [15] that the lattice is affected by the onset of superconductivity and that it contributes substantially to it.

In the following we start from the antiferromagnetic parent compounds of HTSC. The relevant first component is attributed to the $CuO_2$ planes which can be cast into a pure t-J model. Doping has very dramatic consequences since all energy scales are destabilized. Since the large Hubbard U at the copper site prevents the holes from occupying d-orbitals, in a scenario that neglects Cu - O covalency, the oxygen ion p-orbitals will be occupied by the doped holes. As a consequence and to achieve a low energy state, the hole spin aligns antiparallel with respect to the nearest neighbour copper to form a spin singlet state [16]. In addition hole doping induces a strong electron-phonon coupling, particularly around the ($\pi$,0) point of the in-plane high-energy LO phonon [17, 18, 19]. This phonon induces a $Q_2$-type Jahn-Teller distortion, which from symmetry consideration, is the only candidate to strongly modify the orbitals involved and to produce charge transfer between Cu and O. Indeed a marked anomaly that changes with temperature was observed for this mode by neutron inelastic scattering [5, 6]. Even though it has frequently been argued that this singlet state causes HTSC, we have shown recently [20] that this state is rather localized due to the antiferromagnetic background and the strong coupling to phonons. The antiferromagnetic background prevents direct nearest neighbour hopping of the singlet since this can only take place through energetically unfavourable triplet formation or spin flip. The coupling to the $Q_2$ mode leads to an exponential reduction of the hopping matrix element and thus even more localizes this state. The conclusion from the above considerations is that the spin related in-plane channel provides a stable low energy state, but mobility is achieved only via second order processes.

The second component in this scenario relates to the Cu $d_{3z^2-r^2}$ and O $p_z$ orbitals perpendicular to the planes. Here also a strong coupling to the phonons takes place and is most effective for the polar "ferroelectric" low energy mode. The close structural analogy to ferroelectric perovskite oxides suggests that a similar instability could occur in HTSC leading to the formation of a charge density wave instability. The orbitals oriented in-plane and along the c-axis, respectively, are orthogonal and consequently do not interact without doping. Specifically, in a highly symmetric structure with a flat $CuO_2$ plane, the plane and the c-axis components of phonons and orbitals are orthogonal and consequently do not interact. However, when the structure is distorted this orthogonality is violated and they start to interact. For instance, if the Cu-O-Cu bond-angle deviates from 180° by buckling or tilting, the in-plane and c-axis phonon modes become coupled. These distortions occur due to changes in the chemical bonding by doping, and play an important role as we will discuss below. Such distortions are usually local, rather than collective, and can be static as well as dynamic [15, 21].

In order to describe this coupling new hopping elements have to be introduced which admit for processes like hopping from $d_{x^2-y^2}$ - $p_z$, $p_z$ - $d_{3z^2-r^2}$, $d_{3z^2-r^2}$ - $p_x$, $p_y$ etc. In addition strongly anharmonic interactions between in-plane and c-axis phonons take place [22]. It is important to note here that these anharmonic interaction terms give rise to spatial modulations of the ionic displacement coordinates which in turn can induce nanoscale inhomogeneity. Assuming for simplicity that the in-plane states can be described by a single already strongly p - d hybridized band and making the same simplification for the c-axis, this two-component system can be modelled by the following Hamiltonian:

$$H = \sum_{i,\sigma} E_{xy,i} c^+_{xy,i,\sigma} c_{xy,i,\sigma} + \sum_{j,\sigma} E_{z,j} c^+_{z,j,\sigma} c_{z,j,\sigma} + \sum_{i,j,\sigma,\sigma'} T_{xy,z}[c^+_{xy,i,\sigma} c_{z,j,\sigma'} + h.c.] + \sum_{i,j} \tilde{T}_{xy} n_{xy,i\uparrow} n_{xy,j\downarrow}$$
$$+ \sum_{i,j,\sigma,\sigma'} \tilde{V}_C n_{i,\sigma} n_{j,\sigma'} + \sum_{i,j,\sigma,\sigma'} V_{pd} n_{i,\sigma} n_{j,\sigma'}$$

(1)

Here $c^+c=n$ is the plane ($xy$), c-axis ($z$) electron density at site $i,j$ with energy $E$ and spin index $\sigma$. $T_{xy,z}$ is the hopping integral between plane and c-axis orbitals, and $\tilde{T}_{xy}$ is the in-plane spin singlet hopping integral from which a d-wave symmetry of a superconducting order parameter would result. $V_C$ as well as $V_{pd}$ are density-density interaction terms referring to plane / c-axis and in-plane elements. The phonon contributions have already been incorporated in equation 1, where all energies given are renormalized quantities [22, 23]:

$$E_{xy,i} = [\varepsilon_{xy,i} - (g_i^{(xy)} Q_l^{(xy)} - \tilde{g}_{i,m}^{(xy,z)} Q_m^{(z)} <n_{z,m}>)] =$$
$$[\varepsilon_{xy,i} - \{\Delta_i^{xy} + f(xy,z)\}]$$

$$E_{z,j} = [\varepsilon_{z,j} - (g_j^{(z)} Q_m^{(z)} - \tilde{g}_{j,l}^{(xy,z)} Q_l^{(xy)} <n_{xy,l}>)] =$$
$$[\varepsilon_{z,j} - \{\Delta_j^z + f(z,xy)\}]$$

$$T_{xy,z} = [t_{xy,z} - (\tilde{g}_{i,j}^{(xy,z)} \sqrt{Q_l^{(xy)} Q_m^{(z)}})] \exp(-\Phi_T) =$$
$$[t_{xy,z} - (\Delta_{i,j}^{xy;z})] \exp(-\Phi_T^{(xy,z)})$$

$$\tilde{T}_{xy} = [(t_{pd}^2 \frac{\tilde{U}_d - (E_{xy,i} - E_{z,j})}{(E_{xy,i} - E_{z,j})[\tilde{U}_d - (E_{xy,i} - E_{z,j})]}) -$$
$$(\hat{g}_{i,i+2}^{(xy)}(Q_l^{(xy)} + Q_{l+2}^{(xy)})^2] \exp(-\Phi_T^{(xy)})$$

$$\tilde{U}_d = U_d - \frac{(g_i^{(xy)})^2}{m_{xy}\omega_{xy}^2}; \quad \tilde{V}_c = V_c - \frac{(\tilde{g}_{ij}^{xy,z})^2}{\sqrt{m_{xy}m_z}\omega_{xy}\omega_z}$$

$$V_{pd} = \hat{g}_i^{(xy)} Q_l^{(xy)} + \tilde{g}_{i,j}^{(xy,z)}(Q_l^{(xy)} + Q_m^{(z)})$$

(2)

In equation 2 the $\varepsilon$ are the unrenormalized band energies, $g, \tilde{g}$ are the intraband and interband electron-phonon couplings, respectively, and. $Q_l$ is the site $l$ dependent ionic displacement coordinate. The electron-phonon coupling $\hat{g}$ accounts for the symmetry of the $Q_2$ mode and the fact that the singlet state cannot hop to nearest neighbour sites, thus avoiding spin flips or triplet formation. $\omega_{xy}=42.1 meV$ is the in-plane $Q_2$ mode and $\omega_z=22.3 meV$ refers to the ferroelectric type c-axis mode. Both electron correlation terms are locally renormalized with a strong suppression of the $U_d$ – term in each second cell and two-phonon modulation proportional to buckling / tilting. Similarly the Coulomb correlations $V_c$ are reduced and additional phonon driven density-density interactions (proportional to $V_{pd}$) appear which lead to its further reduction. This hybridization term is a consequence of electron-phonon interactions only and favours the hopping between in-plane and out-of-plane orbitals. Another important observation is the exponential supression of the singlet related hopping integral, where $\Phi_T^{(xy,z)} = 1/2N \sum f(q) \frac{\tilde{g}_{i,j}^{(xy,z)}}{\hbar} \sqrt{\frac{Q_l^{(xy)} Q_m^{(z)}}{\omega_{xy}\omega_z}}$, with $f(q)$ a function of the scattering angle, and a corresponding expression for $\Phi_T^{(xy)}$, which induces the rapid suppression of antiferromagnetism with doping. [24]. From the structure of equation 2 it becomes clear that two instabilities should be observed in this system, which is in agreement with recent experimental findings [25]: In the charge channel a CDW instability could set in while in the spin channel a transition to an SDW state could occur. A stabilization of both these instabilities is obtained through the coupling between the components and through anharmonic terms, which have been omitted here for simplicity but have to be explicitly included in realistic modelling [22]. In analogy with ref.26, the two corresponding transition temperatures can be calculated. The higher (charge instability) temperature is identified with the onset temperature of stripe (i. e. charge / lattice inhomogeneity) formation T*. In figure 1 T* is shown as function of the phonon induced electronic gap proportional to $\Delta_j^z = \Delta^* = g_j^{(z)} Q_m^{(z)}$ (the definition of $\Delta_i^{xy}$ is equivalent). Since our emphasis here is not on the isotope effect on T* [1, 22, 27] which is a consequence of the buckling / tilting induced anharmonicity, but on the consequences of the striped phase on superconducitivty, equation 1 is recast into an effective BCS Hamiltonian [28]:

$$H_{eff} = \sum_{i,j,\sigma} \varepsilon_{xy,i}[1 - \Delta_i^{xy} + f(z,xy)]c_{xy,i,\sigma}^+ c_{xy,i,\sigma}$$
$$+ \sum_{i,j,\sigma} \varepsilon_{z,j}[1 - \Delta_j^z + f(xy,z)]c_{z,j,\sigma}^+ c_{z,j,\sigma}$$
$$+ \sum_{i,m} V_{im} c_{xy,i\uparrow}^+ c_{xy,m\downarrow}^+ c_{xy,i\downarrow} c_{xy,m\uparrow} + \sum_{j,n} V_{jn} c_{z,j\uparrow}^+ c_{z,n\downarrow}^+ c_{z,j\downarrow} c_{z,n\uparrow}$$
$$+ \sum_{i,j,m,n} V_{ijmn}[c_{xy,i+j\uparrow}^+ c_{xy,n+m\downarrow}^+ c_{z,i+j\downarrow} c_{z,n+m\uparrow} + c_{z,i+j\uparrow}^+ c_{z,m+n\downarrow}^+ c_{xy,i+j\downarrow} c_{xy,m+n\uparrow}]$$

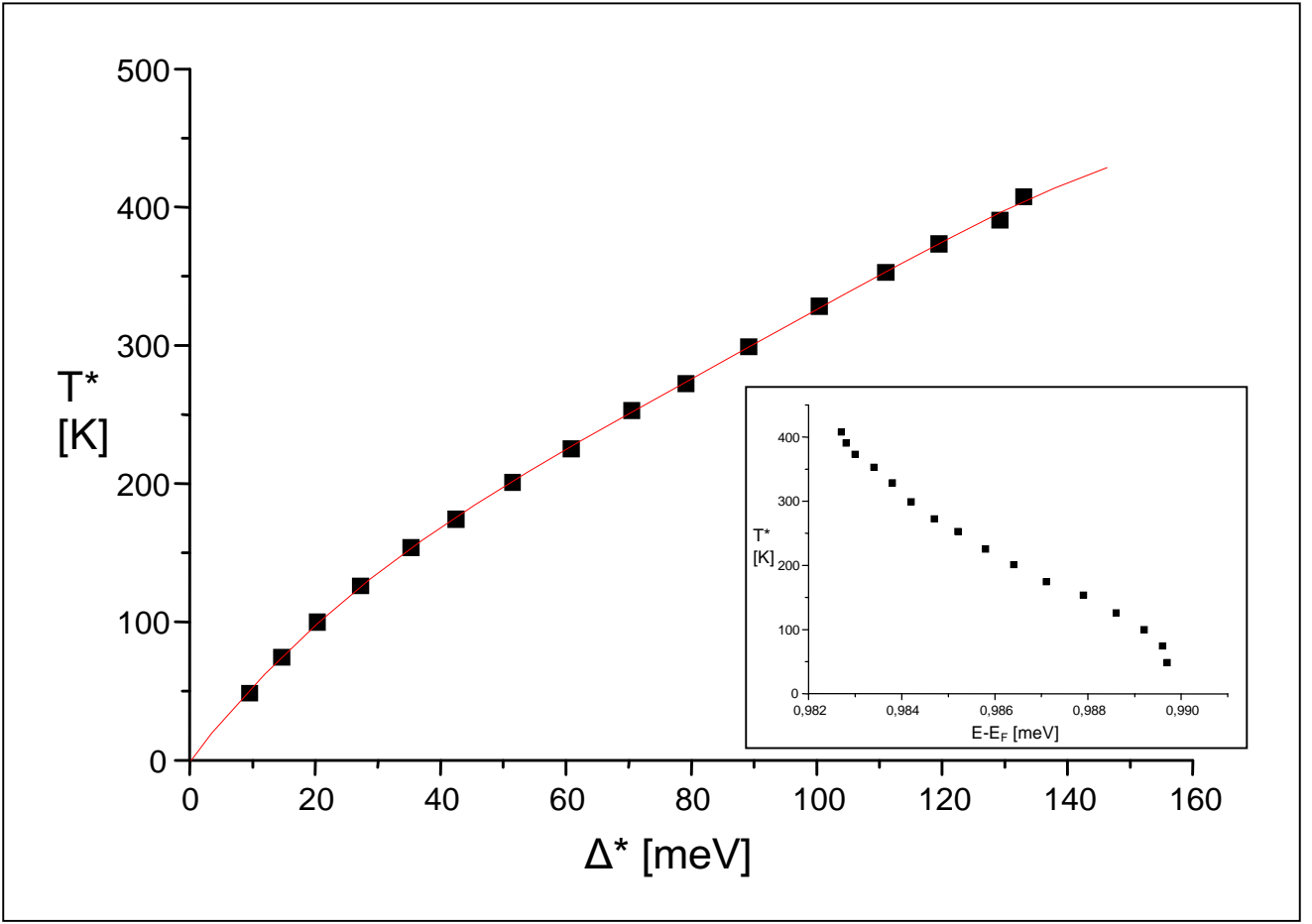

Fig. 1 T* plotted as a function of the phonon induced gap proportional to Δ*. The insert shows T* as a function of doping proportional to E-E$_F$.

(3),

where in the last three terms of euq.3 $i, m, j, n$ denote momenta $k_{xy}, k'_{xy}, k_z, k'_z$. The effects of the Coulomb interactions are included in the effective interaction constants $V$. It is most important that the spin / charge related gaps act on the single particle site energies. The mixing of both components is not only through the effective interaction $V_{ijmn}$, which stems from buckling / tilting, but also through the site energies where all mixing terms have been accommodated in $f$ (see equ. 2). Equation 3 can be solved through a standard Bogoliubov transformation to yield the gap equations as well as the corresponding superconducting transition temperature $T_c$. We start with the assumption that the two components when uncoupled are not superconducting, i.e. $V_{in}$, $V_{jm}=0.01$ are too small to create a paired state. With relatively small interband coupling, $V_{ijmn}=0.1$, the system remains nonsuperconducting. When including the effect of the T* related gap $\Delta^*$, superconductivity appears already at small values of $\Delta^*$, but the corresponding $T_c$ remains rather small. Keeping $V_{in}$, $V_{jm}$ unchanged, but increasing the interband interaction $V_{ijmn}$ (see figure 2) and incorporating the effect of $\Delta^*$ yields a very rapid increase of $T_c$ which readily approaches the experimentally observed values. It is important to note here that the in-plane spin related channel will lead to a d-wave superconducting order parameter while the charge channel corresponds to an s-wave superconducting state. In our approach both contributions are mixed, which is in agreement with several experimental findings [25]. It is also important to note that the atomic structure plays a pivotal role in justifying large values of the interband interaction $V_{ijmn}$. As we mentioned above, if the Cu-O-Cu bond angle $\Phi$ is 180° there is no interband coupling. If $\Phi$ is smaller than 180° contraction of a Cu-O bond changes $\Phi$, thus introducing a strong coupling between the in-plane LO mode and the c-axis transverse mode, with the coupling constant proportional to $1/\sin(\pi-\Phi)$. Thus the coupling is strongest when $\Phi$ is very close to 180°, and if the value of $\Phi$ becomes smaller, $T_c$ is expected to decrease. Indeed, that is exactly what has been observed experimentally [29].

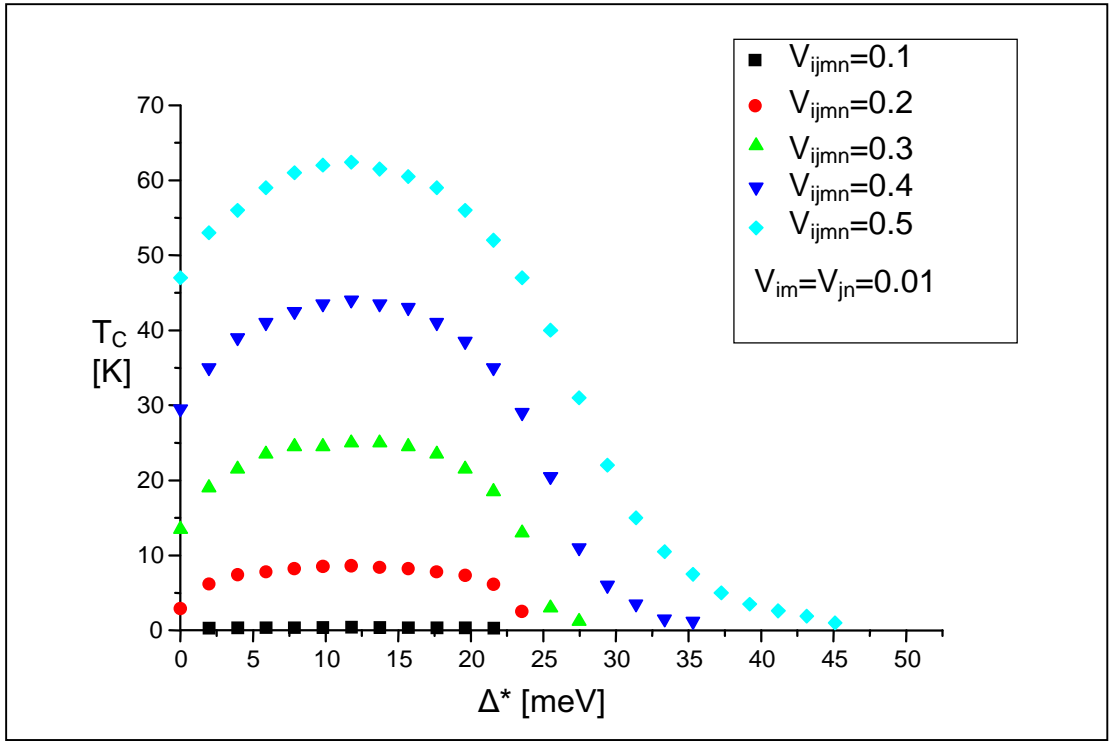

<u>Fig. 2</u> The superconducting transition temperature $T_c$ as a function of $\Delta^*$ for different values of intercomponent couplings $V_{ijmn}$.

In summary, we have studied a very physical and specific two-component scenario to model high temperature superconducting cuprate oxides. The two components are the p-d hybridized bands $d_{x^2-y^2}$ - $p_x$, $p_y$ and $d_{3z^2-r^2}$ - $p_z$. Both components couple strongly to the lattice and experience renormalizations which could induce an SDW or a CDW instability. Doping has the important additional effect of inducing buckling / tilting and strong anharmonicity which allows for coupled plane / c-axis hopping processes. The stripe formation is a consequence of this lattice response and provides the glue to couple the two components. The stripe induced gap brings the corresponding energy levels into resonance and induces a strong enhancement of the superconducting transition temperature, even if the uncoupled components are nonsuperconducting. In conclusion, we find that the doping and lattice induced inhomogeneous nanoscale phase separation provides for spin / charge mixing and that superconductivity is a consequence.

Acknowledgement: One of us (K. A. M.) should like to thank J. Mesot for bringing our attentions on to Ref. 4. A. B.-H. wants to acknowledge the comments by Guo-meng Zhao on the figures which contained errorneous results.